\documentclass[12pt,a4paper]{article}
\usepackage[T1]{fontenc}
\usepackage{dfttrob}
\usepackage{latexuseful2e}
\usepackage{amsmath}
\usepackage{graphicx}

\dfttnum{FISIST/11-2000/CENTRA\\August 2000\\(revised)}

\newcommand{\etaNANA}{\eta_{{}_{N_AN_A}}}
\newcommand{\nuNA}{\nu_{{}_{N_A}}}

\begin{document}

\title{Particle Densities in Heavy Ion Collisions at High Energy
 and the Dual String Model}
\author{J. Dias de Deus and R. Ugoccioni\\
 \it CENTRA and Departamento de F{\'\i}sica (I.S.T.),\\ 
 \it Av. Rovisco Pais, 1049-001 Lisboa, Portugal}
\maketitle

\begin{abstract}
We analyse recent results on charged particle pseudo-rapidity
densities from RHIC in the framework of the Dual String Model, in
particular when including string fusion.
The model, in a simple way, agrees with all the existing data and is
consistent with the presence of the percolation transition to the
Quark-Gluon Plasma already at the CERN-SPS.
\end{abstract}


Recent results on charged particle pseudo-rapidity densities in
central Au+Au collisions, at $\sqrt{s} = 56$ and $\sqrt{s} = 130$
AGeV, presented by the PHOBOS Collaboration, at RHIC, \cite{phobos:1},
give very interesting information that may help to clarify the way the
expected Quark-Gluon Plasma (QGP) is approached as the energy
increases.
Those data also allow to select among different models of particle
production.
As in this experiment the charged particle densities and the average
number of participating nucleons are simultaneously measured, that
provides additional strong constraints to models.

As nuclei are made up of nucleons, it is natural to start by building
nucleus-nucleus collisions as resulting from superposition of
\emph{nucleon-nucleon} collisions, in the way it is done in the
Glauber model approach and generalisations of it.
In one (low energy) limit the nucleons are seen as structureless and
emit particles only in their first collision: this is the wounded nucleon
model \cite{WNM}.
The prediction for particle density, when $N_A$ nucleons from each one of
the nuclei in a AA collision participate, is
\begin{equation}
	\left.\frac{dN}{dy}\right|_{N_AN_A} = 
		\left.\frac{dN}{dy}\right|_{pp} N_A  ,
		\label{eq:wnm_limit}
\end{equation}
where $dN/dy$ is the particle rapidity (or pseudo-rapidity) 
density (for $N_AN_A$ and nucleon-nucleon collisions).
If the nucleon is seen as made up of quarks and gluons, with a growing
number of participating sea quarks and gluons as the energy increases,
one anticipates dominance of multi-collision processes 
\cite{components:AA} and the relation
\begin{equation}
	\left.\frac{dN}{dy}\right|_{N_AN_A} = 
		\left.\frac{dN}{dy}\right|_{pp} \nuNA  ,
		\label{eq:upper_limit}
\end{equation}
to hold, where $\nuNA$ is the number of nucleon-nucleon collisions
when $N_A$ nucleons participate.
Elementary multi-scattering arguments \cite{Armesto:SFM} give
\begin{equation}
	\nuNA = N_A^{4/3}							\label{eq:nu_NA}
\end{equation}

\begin{figure}
  \begin{center}
  \mbox{\includegraphics[width=0.7\textwidth]{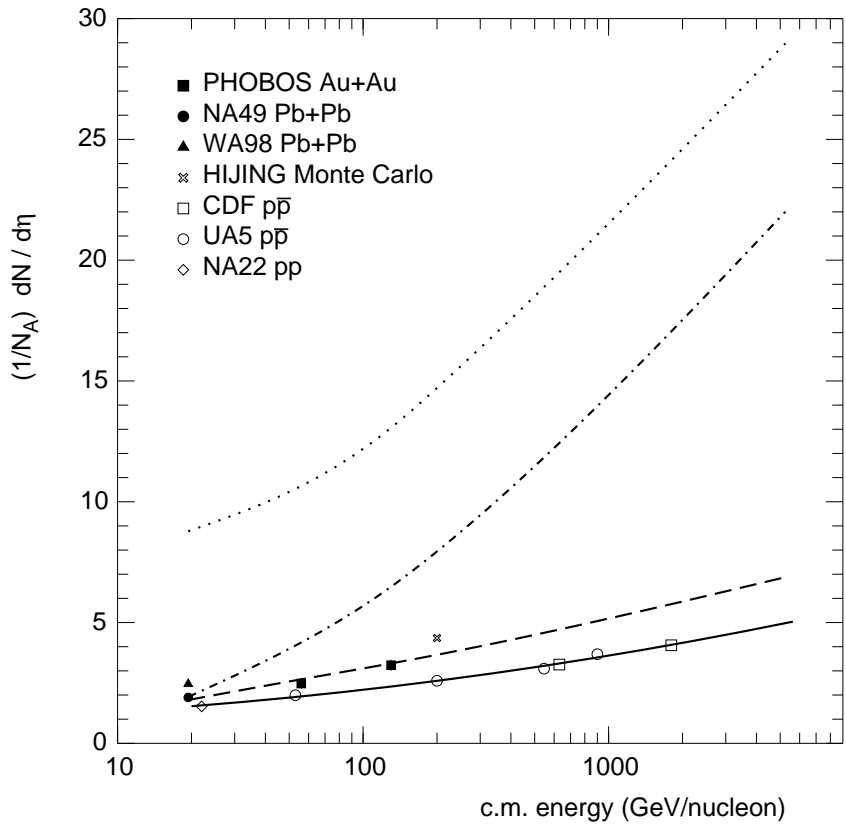}}
  \end{center}
  \caption{Pseudo-rapidity density normalised per participant pair
  as a function of c.m.\ energy. The lines give predictions for the
  wounded nucleon model Eq.~(\ref{eq:wnm_limit}) (solid line), 
  the pure multicollision approach Eq.~(\ref{eq:upper_limit})
  (dotted line), and the Dual String Model, without fusion 
  Eq.~(\ref{eq:nofusion})
  (dash-dotted line) and with fusion Eq.~(\ref{eq:fusion})
  (dashed line). AA points are taken from \cite{phobos:1,WA98:1,Wang},
	$pp$ and $p\bar p$ from \cite{NA22:b4,UA5:rep,UA5:3,CDF:dNdeta}
  }\label{fig:dNdeta}
  \end{figure}

In Fig.~\ref{fig:dNdeta}, together with the PHOBOS data, 
we have presented the quantity
$\frac{1}{N_A} \left.\frac{dN}{dy}\right|_{N_AN_A}$ as function of the
c.m.\ energy $\sqrt{s}$ for the bounds (\ref{eq:wnm_limit})---solid
line---and (\ref{eq:upper_limit}) with (\ref{eq:nu_NA}) ---dotted
line.
We used for $\left.{dN}/{dy}\right|_{pp}$ the 
parametrisation $0.957 + 0.0458\ln(\sqrt{s}) + 0.0494\ln^2(\sqrt{s})$,
with $\sqrt{s}$ in GeV,
which fits data from $pp$ and $p\bar p$ non-single-diffractive
collisions for c.m.\ energies $\sqrt{s} \geq 22$ GeV. The
parametrisation used in \cite{phobos:1,CDF:dNdeta} could not be used
here because it does not fit NA22 data.

\begin{figure}
  \begin{center}
  \mbox{\includegraphics[width=0.9\textwidth]{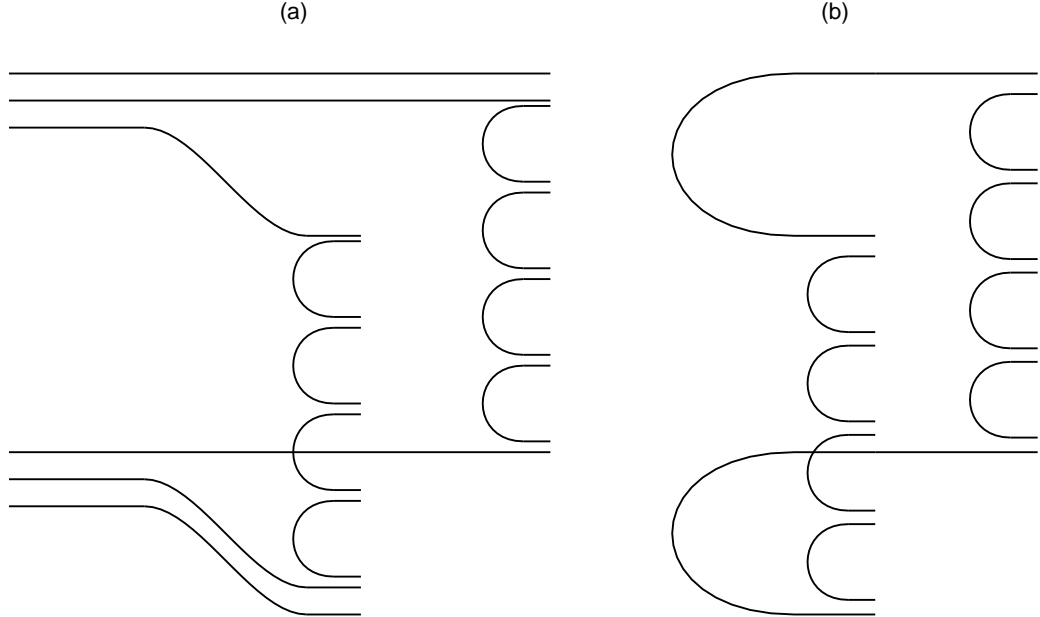}}
  \end{center}
  \caption{Two strings are produced by each
  collisions of valence partons (a) and sea partons (b). Notice how
  the nucleon is broken up in (a), so that further interactions are of
  the type illustrated by (b). See the text for further 
  discussion.}\label{fig:strings}
  \end{figure}

In the Dual String Model (DSM), i.e., the Dual Parton Model \cite{DPM:1}
with the inclusion of strings \cite{StringFusionModel}, the limits 
referred to above appear in a natural way.
The valence quarks of the nucleon produce particles, via strings, only
once ---this is the wounded nucleon model case--- and production is
proportional to the number $N_A$ of participant 
nucleons (Fig~\ref{fig:strings}a).
As the energy and $N_A$ increase the role of sea quarks and gluons
increases, they interact and produce, again via strings, particles, and
the number of collisions $\nu$ becomes the relevant 
parameter (Fig~\ref{fig:strings}b).

One should notice that the diagram of Fig.~\ref{fig:strings}b
may be interpreted as
multiple inelastic scattering, either internally within a given
nucleon-nucleon collision or externally involving interactions with
different nucleons.
On the other hand, this diagram may appear repeated several times.

Following \cite{Armesto:SFM}, and taking into account the basic
diagrams of Fig.s~\ref{fig:strings}a and \ref{fig:strings}b, we now 
write an expression for the particle pseudo-rapidity density,
\begin{equation}
	\left.\frac{dN}{dy}\right|_{N_AN_A} = 
		N_A \left[ 2 + (2k-1)\alpha\right] h
				+ (\nuNA - N_A)  2k\alpha h ,	
																			\label{eq:Armesto}
\end{equation}
where $h$ is the height of the valence-valence rapidity plateau,
$\alpha$ is the relative weight of the sea-sea (including gluons)
plateau and $k$ is the average number of string pairs per collision.
The diagrams of Fig.s~\ref{fig:strings}a and \ref{fig:strings}b
correspond to $k=1$.
However, as we mentioned above, the diagram of Fig.~\ref{fig:strings}b
can be iterated with $k\geq 1$ being, in general, a function of energy.
The number of nucleon-nucleon collisions is, of course,
\begin{equation}
	N_A + (\nuNA - N_A) = \nuNA   ,
\end{equation}
and the number $N_s$ of strings is
\begin{equation}
	N_s = N_A\left[ 2 +2(k-1)\right] + (\nuNA - N_A) 2k =
			2k\nuNA  .
\end{equation}

The first term on the right-hand side of Eq.~(\ref{eq:Armesto})
is just a sum over
nucleon-nucleon scattering contributions (including internal parton
multiple scattering) and we can thus write
\begin{equation}
		\left.\frac{dN}{dy}\right|_{N_AN_A} = \left.\frac{dN}{dy}\right|_{pp}
				N_A + (\nuNA - N_A) 2k\alpha h  ,
																\label{eq:nofusion}
\end{equation}
with
\begin{equation}
		\left.\frac{dN}{dy}\right|_{pp} = \left[ 2 +2(k-1)\alpha \right] h   .
																\label{eq:pp}
\end{equation}
If external multiple scattering is absent, by putting $\nuNA =
N_A$, one obtains the wounded nucleon model limit, Eq.~(\ref{eq:wnm_limit}).
If multiple scattering dominates, $k \gg 1$, we obtain the limit of
Eq.~(\ref{eq:upper_limit}).

In order to make more transparent the comparison with data, we shall
rewrite Eq.~(\ref{eq:nofusion}), by using Eq.~(\ref{eq:pp}) and
Eq.~(\ref{eq:nu_NA}), in the form
\begin{equation}
	\frac{1}{N_A} \left.\frac{dN}{dy}\right|_{N_AN_A} = 
		\left.\frac{dN}{dy}\right|_{pp} N_A^{1/3} 
				-  (N_A^{1/3} -1) 2 (1-\alpha) h   .			\label{eq:nofusion:bis}
\end{equation}
We show the result of this model in Fig.~\ref{fig:dNdeta}
(dash-dotted line). From comparison of Eq.~(\ref{eq:pp}) with $pp$
data at low energy, $k\simeq 1$, one obtains $h \simeq 0.75$.
The parameter $\alpha$ in Eq.~(\ref{eq:nofusion:bis}) was put equal to
0.05.

In the Dual String Model the strings interact, the simplest
interaction being fusion due to overlap in the transverse 
plane \cite{StringFusionModel}.
This is the mechanism that leads to percolation and to the
Quark-Gluon Plasma formation \cite{Pajares:perc,NardiSatz,jotapsi}.
When strings fuse, the strength of the colour field is reduced in
comparison with the colour field generated by the same number of
independent strings. This is essentially due to the random sum of
colour vectors \cite{Biro:randomsum}: $Q_n^2 = \sum_{i=1}^{n} Q_i^2$
and $Q_n = \sqrt{n} Q$ if all the $n$ strings are of the same type.

Introducing the dimensionless transverse density percolation parameter
$\eta$,
\begin{equation}
	\eta \equiv \frac{r_s^2 N_s}{R_{N_A}^2}  ,
\end{equation}
where $r_s$ is the string transverse radius (we shall take $r_s = 0.2$
fm, see \cite{Pajares:perc,Satz:98}), $R_{N_A}$ the radius of the 
interaction area ($R_{N_A}\simeq 1.14 N_A^{1/3}$) and $N_s$ the number
of strings, the effective reduction factor in particle production
is \cite{Braun:Feta},
\begin{equation}
	F(\eta) = \sqrt{ \frac{1-e^{-\eta}}{\eta} }   .
\end{equation}
As $\eta\to 0$, $F(\eta)\to 1$ (no fusion) and as $\eta\to\infty$,
$F(\eta)\to 1/\sqrt{\eta} \approx 1/\sqrt{N_s}$ (all the strings fuse).

We can consider the parameter $\eta$ in two situations.
In nucleon-nucleon internal interactions, and then
\begin{equation}
	\eta_{pp} \equiv \frac{r_s^2}{R_{pp}^2}  [2 +(2k-1)] = 
			\frac{r_s^2}{R_{pp}^2} 2k   .
																	\label{eq:eta_pp}
\end{equation}
At present energies $\eta_{pp}$ is negligible, $\eta_{pp} \approx
10^{-2}\div 10^{-1}$. But we can also consider $\eta$ in external interactions,
with
\begin{equation}
	\etaNANA = \frac{r_s^2}{R_{N_A}^2} 2k(\nuNA - N_A) 
			\simeq \left(\frac{r_s}{1.14}\right)^2 2k (N_A^{1/3} - 1) N_A^{1/3} .
																\label{eq:eta_AA}
\end{equation}
For $N_A \approx 10^2$, as in \cite{phobos:1}, $\etaNANA > 10
\eta_{pp}$ and we shall then only consider $\etaNANA$.

Eq.~(\ref{eq:Armesto}) with string fusion becomes
\begin{equation}
	\begin{split}
	\frac{1}{N_A} \left.\frac{dN}{dy}\right|_{N_AN_A}  
	& = 
		\left.\frac{dN}{dy}\right|_{pp} \left[1-F(\etaNANA)\right]\\
	& +
		F(\etaNANA) \left[
		\left.\frac{dN}{dy}\right|_{pp} N_A^{1/3}
				-  (N_A^{1/3} -1) 2 (1-\alpha) h \right]  .
	\end{split}
																\label{eq:fusion}	
\end{equation}
In Fig.~\ref{fig:dNdeta} we have also shown the prediction of the DSM
with string fusion (dashed line) again with $h=0.75$ and
$\alpha=0.05$.  The deviation from the wounded
nucleon model limit becomes weaker and the agreement with PHOBOS data
is quite satisfactory.

We would like now to make a few comments:

1. The predictions for particle densities in central Pb+Pb collisions
   of the DSM without fusion and of the DSM with fusion are very
   different at $\sqrt{s} = 200$ AGeV (RHIC) and at $\sqrt{s} = 5.5$
   ATeV (LHC) as can be seen in the following table, showing the
   average pseudo-rapidity density in the interval $[-1,1]$:
	\begin{center}
  \begin{tabular}{c|c|c}
	\hline
	c.m.\ energy   & 200 AGeV & 5.5 ATeV \\
	\hline
	without fusion & 1500     &  4400    \\
	with fusion    &  700     &  1400    \\
	\hline
  \end{tabular}
	\end{center}

2. The models considered here are essentially soft models. The
   parameters of the elementary collision densities, $h$ and $\alpha$,
   were assumed constant, all the energy dependence being attributed
   to the parameter $k$, the average number of string pairs per
   elementary collision.
	 If $h$ and $\alpha$ are allowed to grow with energy, as a result,
   for instance, of semi-hard effects, the parameter $k$ may then have
   a slower increase than the one obtained here.

3. The value found for $\alpha$, $\alpha\simeq 0.05$, means that the
   height of the sea-sea plateau is much smaller than the height of
   the valence-valence plateau. By noticing that for valence-valence
   collisions the two strings stretch all over forward/backward
   rapidity without much overlap, while for sea-sea collisions the two
   strings do overlap, the value found for $\alpha$ means
	\begin{equation}
			\left.\frac{dN}{dy}\right|_{\textrm{sea-sea}} \simeq
				0.1  \left.\frac{dN}{dy}\right|_{\textrm{val-val}}  .
	\end{equation}

4. In our Dual String Model with fusion, the parameter
	 $\etaNANA$ at the CERN-SPS
   has the value $\etaNANA \approx 1.8$, larger than the critical
   density ($\eta_c \approx 1.12\div1.17$) which means that percolation
   transition is already taking place at $\sqrt{s} = 20$ AGeV, even
   allowing for non-uniform matter distribution in the nucleus
   ($\eta_c\approx 1.5$) \cite{perc:98}; this result is valid
	 even with $k=1$. 
	 The observed anomalous $J/\psi$ suppression \cite{NA50:QGP} may
   then be a signature of the percolation transition to the
   Quark-Gluon Plasma \cite{jotapsi}.

\bigskip
After the submission of this paper, we became aware of two
papers on the same subject \cite{Wang,WA98:1}.

1. From the paper of Wang and Gyulassy 
   \cite{Wang} we realised that the HIJING point at 200 AGeV in
 	 Ref.~\cite{phobos:1} was 20\% too high. This point was corrected in our
   figure. 

2. In the WA98 Collaboration paper on scaling of particle and
   transverse energy
   production \cite{WA98:1}, results on $dN/dy$ were presented 
	 in Pb+Pb collisions at 158 A GeV. This point was included in
   Fig.~\ref{fig:dNdeta} but not taken into account in the
   calculations. It somewhat disagrees with the NA49 point presented
   in \cite{phobos:1}.

\section*{Acknowledgements}
We thank X.-N. Wang and U. Heinz for comments on recent work.
R.U. gratefully acknowledges the financial support of the
Fundação Ciência e Tecnologia via the ``Sub-Programa Ci\^encia 
e Tecnologia do $2^o$ Quadro Comunit\'ario de Apoio.''

\newpage
\section*{References}
\bibliographystyle{prstyR}  
\bibliography{abbrevs,bibliography}

\end{document}